\documentclass[sigconf]{acmart}

\AtBeginDocument{%
  \providecommand\BibTeX{{%
    \normalfont B\kern-0.5em{\scshape i\kern-0.25em b}\kern-0.8em\TeX}}}

\copyrightyear{2021}
\acmYear{2021}
\setcopyright{acmcopyright}\acmConference[SIGIR '21]{Proceedings of the 44th International ACM SIGIR Conference on Research and Development in Information Retrieval}{July 11--15, 2021}{Virtual Event, Canada}
\acmBooktitle{Proceedings of the 44th International ACM SIGIR Conference on Research and Development in Information Retrieval (SIGIR '21), July 11--15, 2021, Virtual Event, Canada}
\acmPrice{15.00}
\acmDOI{10.1145/3404835.3462895}
\acmISBN{978-1-4503-8037-9/21/07}

\usepackage{booktabs}
\usepackage{multirow}
\usepackage{algorithm}
\usepackage{algorithmic}
\usepackage{enumitem}
\usepackage{xcolor}

\begin{document}

\title{A Graph-Enhanced Click Model for Web Search}

\author{Jianghao Lin$^1$, Weiwen Liu$^2$, Xinyi Dai$^1$, Weinan Zhang$^1$, Shuai Li$^1$ \\ Ruiming Tang$^2$, Xiuqiang He$^2$, Jianye Hao$^2$, Yong Yu$^1$}
\affiliation{$^1$Shanghai Jiao Tong University, $^2$Huawei Noah's Ark Lab
    \city{}
    \state{}
    \country{}}
\email{{chiangel, daixinyi, wnzhang, shuaili8, yyu}@sjtu.edu.cn, {liuweiwen8, tangruiming, hexiuqiang1, haojianye}@huawei.com}

\renewcommand{\shortauthors}{Jianghao Lin, et al.}

\begin{abstract}

To better exploit search logs and model users' behavior patterns, numerous click models are proposed to extract users' implicit interaction feedback. Most traditional click models are based on the probabilistic graphical model (PGM) framework, which requires manually designed dependencies and may oversimplify user behaviors. Recently, methods based on neural networks are proposed to improve the prediction accuracy of user behaviors by enhancing the expressive ability and allowing flexible dependencies. However, they still suffer from the data sparsity and cold-start problems. In this paper, we propose a novel graph-enhanced click model (GraphCM) for web search.
\emph{Firstly}, we regard each query or document as a vertex, and propose novel homogeneous graph construction methods for queries and documents respectively, to fully exploit both intra-session and inter-session information for the sparsity and cold-start problems.
\emph{Secondly}, following the examination hypothesis\footnote{Examination hypothesis: a user clicks a document if and only if she examines the document and is attracted by the document.}, we separately model the attractiveness estimator and examination predictor to output the attractiveness scores and examination probabilities, where graph neural networks and neighbor interaction techniques are applied to extract the auxiliary information encoded in the pre-constructed homogeneous graphs.
\emph{Finally}, we apply combination functions to integrate examination probabilities and attractiveness scores into click predictions. Extensive experiments conducted on three real-world session datasets show that GraphCM not only outperforms the state-of-art models, but also achieves superior performance in addressing the data sparsity and cold-start problems.

% \emph{Firstly}, we regard each query or document as vertex and build homogeneous graphs for query and document respectively, to exploit cross-session, reformulation and other auxiliary information. \emph{Secondly}, we separately model an examination predictor and an attractiveness estimator for user behavior modeling, where abundant side information encoded in the pre-constructed homogeneous graphs is leveraged to mitigate the data sparsity and cold-start problems. \emph{Finally}, we apply combination functions to integrate the examination probabilities and attractiveness scores into click prediction. Extensive experiments conducted on three real-world session datasets show that GraphCM not only outperforms the state-of-art models, but also achieves superior performance in addressing the data sparsity and cold-start problems.

\end{abstract}






\maketitle

\section{Introduction}

Understanding users' behavior patterns is key to improving the performance of an information retrieval system. In web search, the ability to summarize users' behavior patterns and precisely simulate user interactions allows search engines to better fulfill users' information needs. To this end, numerous \emph{click models} have been proposed to model users' click behaviors. They serve as click simulators in cases where no real users are available or we prefer not to experiment with real users to avoid hurting user experiences~\cite{borisov2016neural}. Besides,
% \shuai{the logic to get the next sentence is not fluent}
click models are also used to estimate the relevance scores for query-document pairs to facilitate document ranking~\cite{chen2020context,wang2013content,dai2021adversarial}.

Earlier click models are based on the \emph{probabilistic graphical model} (PGM) framework~\cite{koller2009probabilistic}, where user behaviors are represented as a sequence of observable and hidden states (e.g., clicks, skips, attractiveness, and examinations)~\cite{chuklin2015click,zhu2010novel}. PGM-based click models often require manually designed dependencies between each binary variable, which is likely to oversimplify and therefore overlook some key aspects of user behaviors.
To better capture users' behavior patterns and allow flexible dependencies, Borisov~\shortcite{borisov2016neural} proposed a neural click model (NCM), which adopts the \emph{distributed vector representation approach} for user behavior representations, instead of binary variables.
% \shuai{this and above sentences can be combined in some way. and 'binary variables' is kind of repetition since you already used 'to better xxx' } 
While NCM only encodes \textbf{query-level} information, the context-aware click model (CACM)~\cite{chen2020context} utilizes complex structures to incorporate interaction effects among different queries within the same session (i.e., \textbf{intra-session} information) and achieves the state-of-art performance among existing click models. However, these click models fail to consider the following two issues.

\begin{figure}[t]
    % \vspace{10pt}
	\centering
	\includegraphics[width=0.475\textwidth]{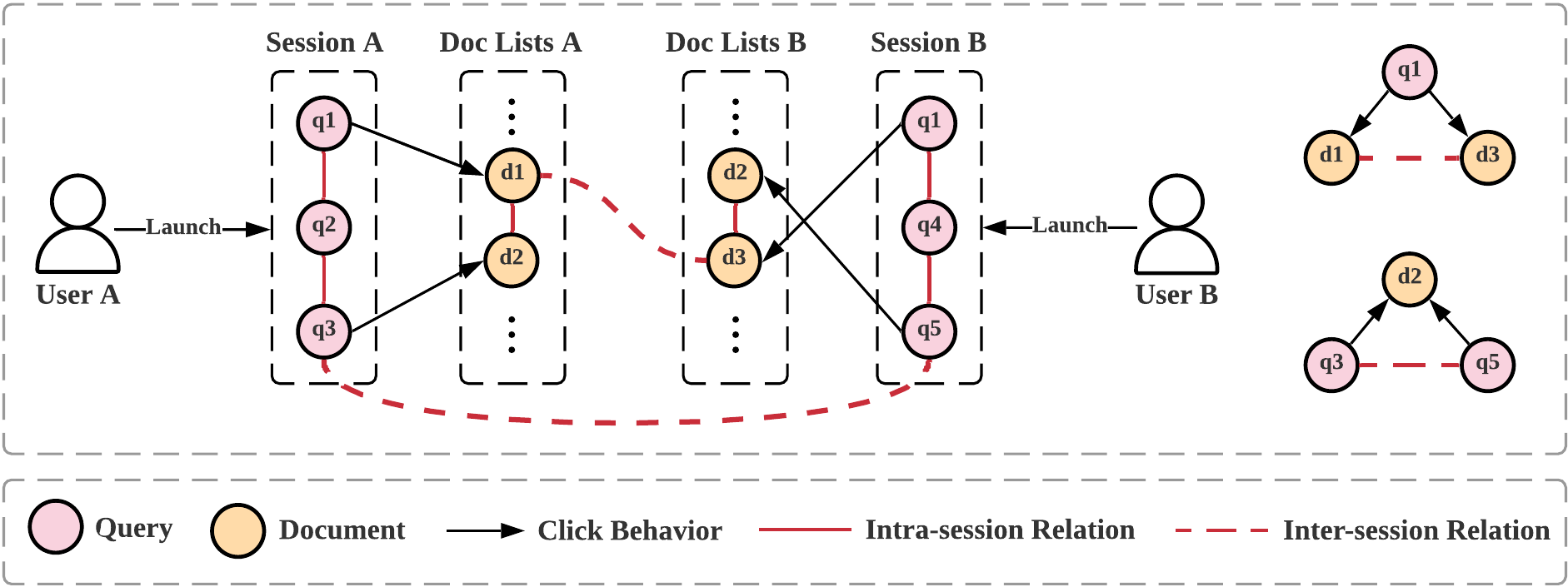}
	\caption{Illustration of intra-session and inter-session information in web search.}
    \vspace{-10pt}
	\label{fig:intra and inter session}
\end{figure}

First, click models generally suffer from the data sparsity problem, i.e.,
% \shuai{rewrite next noun phrase, not very strong} 
the lack of useful user interaction feedback on query-document pairs. Take \emph{TianGong-ST} dataset\footnote{\url{http://www.thuir.cn/tiangong-st/}} for example, the interaction sparsity ratio is $99.9969\%$.\footnote{The interaction sparsity ratio is calculated as $\frac{\#\ \text{missing}\ \text{query-document}\ \text{interactions}}{ \#\  \text{possible}\ \text{query-document}\ \text{pairs}}$.} Such sparsity of user interactions brings difficulty to the training of click models and leads to poor generalization performance.
% \shuai{this statement is not very strong, doesn't emphasize *bottleneck* point of the whole problem, since people can avoid using click models and perhaps these difficulty doesn't matter}
Although incorporating intra-session information as in CACM may alleviate the sparsity problem, it does not fully utilize the user interactions among queries and documents. 
In fact, we find that there is rich potential for extracting users' behavior patterns from interactions between queries or documents across different sessions issued by different users (i.e., \textbf{inter-session} information). As illustrated in Figure~\ref{fig:intra and inter session}, User A and User B launch sessions that consist of different queries towards similar topics
% \shuai{you first says 'similar', but later process with 'may share similar'. a bit strange. need to rewrite this part}
. Then queries from different sessions ($q_3$ and $q_5$) are likely to be related if they positively interact with the same document $d_2$, as they share similar user intents. Likewise, documents from different sessions ($d_1$ and $d_3$) are related as they are clicked by the same query $q_1$. Therefore, by better extracting both intra-session and inter-session information, we can
% \shuai{here is introduction to motivate, we don't have experiments yet. and the motivation shouldn't be *since the result is better*} 
achieve superior performance in mitigating the data sparsity problem.
% Therefore, the data sparsity problem can be mitigated by making use of the collaborative information across different sessions issued by different users. \ww{compared to CACM, we can do better}
%Although CACM has utilized intra-session information, it ignores the possible interaction between queries or documents across different sessions issued by different users (i.e., \textbf{inter-session} information). As shown in Figure~\ref{fig:intra and inter session}, different users may launch sessions that consist of different queries towards similar topics. By the concept of \emph{collaborative filtering} in recommender system, if queries across different sessions click the same document, or documents across different sessions are clicked by the same query, they should provide additional \textbf{inter-session} information for user behavior modeling, which is useful to alleviate the data sparsity problem.

Second, existing click models are vulnerable in the cold-start environments. That is, previous click models generally fail to
% \shuai{these claims are too strong and might hurt authors of previous works. Actually even though some models are not designed for general setting, they can general a bit, depending how general setting in expectation} 
effectively predict user clicks when there are queries or documents in the test set that never appear in the training set, and thus meeting a dramatic performance decrease during the test phase. For example, in Figure~\ref{fig:intra and inter session}, suppose the second query $q_4$ in Session B is a brand-new query launched by User B during the test phase. Existing click models cannot make reliable predictions for $q_4$, since little user behavior information is available to form the basis for predictions.
% \shuai{in most cases, we will not have brand-new information. at least we could have some features possibly of low quality. so be a little careful about this part} 
Actually, as queries in the same session issued by a user share similar intents towards similar topics, adjacent queries can provide useful reformulation information for the inference and enrich the user interaction representations. For instance, $q_4$'s adjacent queries ($q_1$ and $q_5$) in Session B can be regarded as reformulations of $q_4$, and thereby can be leveraged to predict user behaviors.

To tackle the aforementioned limitations, we propose a novel graph-enhanced click model (GraphCM) in this paper.
\emph{Firstly}, we regard each query or document as a vertex and propose novel homogeneous graph construction methods for queries and documents respectively, to fully exploit both intra-session and inter-session information for the sparsity and cold-start problems.
% \emph{Firstly}, we regard each query or document as a vertex and build homogeneous graphs for queries and documents respectively, to exploit inter-session, \ww{reformulation and other side information} for the sparsity and cold-start problems. \ww{flow}
% which contain abundant side information for the sparsity and cold-start problems.
\emph{Secondly}, following the examination hypothesis~\cite{richardson2007predicting}, we separately model the attractiveness estimator and examination predictor to output the attractiveness scores and examination probabilities, where graph neural networks and neighbor interaction techniques are applied to extract the auxiliary information encoded in the pre-constructed homogeneous graphs.
% \emph{Secondly}, we separately model the attractiveness estimator and examination predictor to output the attractiveness scores and examination probabilities. The abundant side information, which is included in the pre-constructed homogeneous graphs, is encoded in the attractiveness estimator via graph neural networks and neighbor interactions.
\emph{Finally}, we combine attractiveness scores and examination probabilities through a combination layer to perform user click prediction.
% \ww{relations between these contributions}
% \shuai{I feel that the flow of this part is not very satisfying. using lists is always a good idea}
Main contributions of this paper are:

\begin{itemize}[leftmargin=10pt]
    \item We propose novel homogeneous graph construction methods for queries and documents respectively, which can fully extract both intra-session and inter-session information.
    \item We propose a graph-enhanced click model (GraphCM) to exploit structural dependencies among queries and documents according to pre-constructed homogeneous graphs. To the best of our knowledge, this is the first work to apply graph neural networks and neighbor interaction techniques
    to the field of click models to address the data sparsity and cold-start problems.
    \item The proposed GraphCM achieves significantly better performance than existing click models in both click prediction and relevance estimation tasks. Besides, extensive empirical studies are conducted to show the ability of GraphCM to tackle the sparsity and cold-start problems.
\end{itemize}

\section{Related Work}
% This work draws on the following research areas: click models and graph representation learning.

\subsection{Click Models}

To model and simulate user behaviors, numerous click models have been proposed for various application scenarios (e.g., web search, recommendation)~\cite{chuklin2015click,zhong2010incorporating,xu2012incorporating,chen2020beyond}. Traditional click models, which are based on the probabilistic graphical model (PGM) framework, treat user behaviors as a sequence of observable and hidden events. They usually incorporate different assumptions on user behaviors to specify how documents and clicks at different positions affect each other.
Most PGM-based click models adopt the examination hypothesis~\cite{richardson2007predicting} where the probability of click are decomposed into the examination probability and the attractiveness score. 
Different click models study the examination probability differently. A simple click model that follows the examination hypothesis is the position-based model (PBM)~\cite{craswell2008experimental}, which assumes that the examination probability is only related to the displayed positions. The cascade model (CM)~\cite{craswell2008experimental} assumes that users scan each document in the list from top to bottom until the first click. Yet CM can only handle query sessions with exactly one click. On the basis of CM, user browsing model (UBM)~\cite{dupret2008user}, dynamic Bayesian network (DBN)~\cite{Chapelle2009DBN}, dependent click model (DCM)~\cite{Guo2009DCM}, and click chain model (CCM)~\cite{Guo2009CCM} have been proposed to overcome this limitation. 

As the dependencies in traditional PGM-based click models are designed manually~\cite{zhang2011user}, some neural network based approaches have been proposed to get better expressive power and flexible dependencies. The neural click model (NCM)~\cite{borisov2016neural} is the first attempt to apply neural networks to click models. NCM treats user behaviors as a sequence of hidden states instead of binary events. The following NN-based click models also adopt this distributed representation framework. The click sequence model (CSM)~\cite{borisov2018click} maintains an encoder-decoder architecture to predict the position sequence of the clicked documents. The context-aware click model (CACM)~\cite{chen2020context} takes the intra-session information into consideration and separately models the examination probability and the attractiveness score. Our proposed GraphCM can better extract both intra-session and inter-session information by applying graph neural networks and neighbor interaction techniques, resulting in the state-of-art performance compared to existing click models.
\vspace{-4pt}
\subsection{Graph Representation Learning}

Graph representation learning investigates low-dimensional representations of graph vertices, while preserving node content and structural information~\cite{zhang2019heterogeneous,ribeiro2017struc2vec,wang2016structural}. Earlier graph representation learning methods learn node representations from the graph structure by applying random walks. DeepWalk~\cite{perozzi2014deepwalk} apply random walks to generate the node context and learn node embeddings. Node2vec~\cite{grover2016node2vec} further exploits a biased random walk strategy to better capture the local and global structural information. In recent years, graph neural networks are proposed to aggregate information from neighbors and encode structural contexts. GCN~\cite{kipf2016semi} applies local graph convolutions for the node classification task. GraphSage~\cite{hamilton2017inductive} performs a non-spectral graph convolution over a fixed size of sampled neighbors to integrate neighbor features for learning accurate node representations. GAT~\cite{kipf2016semi} utilizes a multi-head attention mechanism to increase the model capacity.

\begin{figure*}[t]
	\centering
	\includegraphics[width=\textwidth]{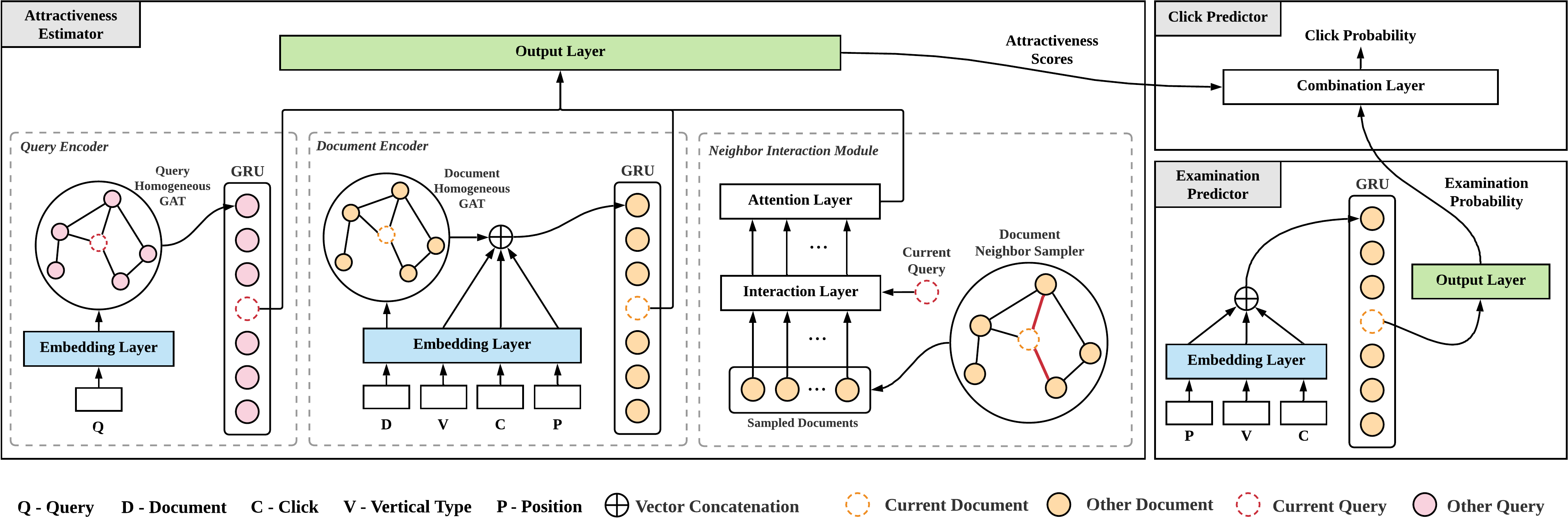}
    \vspace{-13pt}
    \caption{The overall framework of GraphCM. 
     GraphCM consists of an attractiveness estimator and an examination predictor. The attractiveness estimator contains three components (i.e., query encoder, document encoder, and neighbor interaction module), which encode the query context, document context, and query-document interactions to estimate the attractiveness score $\mathcal{A}$. The examination predictor utilizes the session context to predict the examination probability $\mathcal{E}$. GraphCM integrates $\mathcal{A}$ and $\mathcal{E}$ through a combination layer to predict user click behaviors.
    }
    \vspace{-5pt}
    \label{fig:framework}
\end{figure*} 

GCN, GraphSage and GAT are popular architectures of the general graph neural networks, and can be naturally regarded as plug-in graph representation modules for many supervised tasks~\cite{qu2019end}. Recently, researchers also deploy graph neural networks in recommender systems to make full use of structural side information and tackle the data sparsity and cold-start problems. For example, PinSage~\cite{ying2018graph} applies graph neural networks to pin-board graph for recommendation. KGAT~\cite{wang2019kgat} incorporates user-item graph with knowledge graph to generate finer node representations. 
% HetGNN~\cite{zhang2019heterogeneous} jointly considers heterogeneous contents encoding, type-based neighbors aggregation and types combination to perform heterogeneous graph representation learning. 
% \shuai{adding some connections to introduce the basis of our motivation}

Likewise, general graph neural networks are suitable plug-in modules for click models. To the best of our knowledge, we are the first to introduce graph neural networks to the field of click models, which extracts abundant auxiliary information included in the constructed homogeneous graphs. Therefore, our proposed GraphCM can achieve better performance in addressing the data sparsity and cold-start problems.
\vspace{-2pt}
\section{Problem Formulation}
The user browsing behaviors in web search is regarded as a series of independent search sessions. A search session $\mathcal{S}$ can be formulated as a sequence of queries $\mathcal{Q}_N=\left[q_1, ..., q_N\right]$ submitted by the user. For each query $q_i$, the search engine returns a ranked list of documents $\mathcal{D}_{i}=\left[d_{i,1},d_{i,2},...,d_{i,M}\right]$. Each document $d_{i,j}$ has three attributes: the unique URL identifier $u_{i,j}$, the ranking position $p_{i,j}$ and the vertical type\footnote{Vertical type means the presentation style of a displayed document (e.g., organic vertical, the illustrated vertical, the encyclopedia vertical)~\cite{chen2020context}.} $v_{i,j}$. The user browses the ranked lists and may click several documents in the session. We define the click variable $c_{i,j}$ for each document, where $c_{i,j}=1$ if $d_{i,j}$ is clicked by the user and 0 if not. Then we can define the problem of click model tasks as follows:

Given the user's browsing history $\mathcal{Q}=\left[q_1,q_2,...,q_n\right]$, $\mathcal{D}=[d_{1,1},$ $d_{1,2},...,d_{n,m}]$, and $\mathcal{C}=\left[c_{1,1},c_{1,2},...,c_{n,m-1}\right]$, for the m-th document $d_{n,m}$ in the n-th query $q_{n}$ of the session $\mathcal{S}$, we would like to (\romannumeral1) predict whether the document $d_{n,m}$ will be clicked by the user (i.e., the click variable $c_{n,m}$), and (\romannumeral2) estimate the context-aware relevance between $q_n$ and $d_{n,m}$.
\vspace{-2pt}
\section{Model Framework}
In this section, we introduce the framework of graph-enhanced click model (GraphCM), which is shown in Figure~\ref{fig:framework}.
\subsection{Overview of GraphCM}

Most click models adopt the examination hypothesis: a user clicks a document $d_{n,m}$ if and only if she examines the document and is attracted by the document~\cite{richardson2007predicting}, which can be formulated as:
\begin{equation}
    c_{n,m}=1 \Leftrightarrow e_{n,m}=1 \operatorname{and} a_{n,m}=1,
\end{equation}
where $c_{n,m}\in\{0,1\}$, $e_{n,m}\in\{0,1\}$, and $a_{n,m}\in\{0,1\}$ are click, examination, and attractiveness variables, respectively. In this work, for each document $d_{n,m}$, we assume that the user examines it with probability $\mathcal{E}_{n,m}$, and the attractiveness score is $\mathcal{A}_{n,m}$. As shown in Figure~\ref{fig:framework}, GraphCM separately models the attractiveness estimator and examination predictor, and integrates the attractiveness score $\mathcal{A}_{n,m}$ and examination probability $\mathcal{E}_{n,m}$ at click predictor module to output the click probability.
%Variable $e_{n,m}\in\{0,1\}$ denotes whether the user is examined by $d_{n,m}$ with mean $\mathcal{E}_{n,m}$. Variables $\mathcal{E}_{n,m}$ and $\mathcal{A}_{n,m}$ represent the examination probability and attractiveness score of the document $d_{n,m}$. As shown in Figure~\ref{fig:framework}, GraphCM separately models the attractiveness estimator and examination predictor, and integrate the attractiveness score and examination probability at click predictor module to output the click probability.
% \begin{equation}
%     c_{n,m}=1 \Leftrightarrow \mathcal{E}_{n,m}=1 \operatorname{and} \mathcal{A}_{n,m}=1
% \end{equation}
% where $c_{n,m}=1$ denotes that the document $d_{n,m}$ is clicked by the user. Variables $\mathcal{E}_{n,m}$ and $\mathcal{A}_{n,m}$ represent the examination probability and attractiveness score of the document $d_{n,m}$. As shown in Figure~\ref{fig:framework}, GraphCM separately models the attractiveness estimator and examination predictor, and integrate the attractiveness score and examination probability at click predictor module to output the click probability.

Next, we will first introduce the graph construction methods and general embedding layers, and then elaborate on the details of each module (i.e., attractiveness estimator, examination predictor and click predictor).

\subsection{Graph Construction}
\label{sec:graph construction}
Thoroughly incorporating intra-session and inter-session information into user behavior modeling is essential for alleviating the data sparsity and cold-start problems. However, it is challenging to fuse complex user behaviors among queries and documents. Moreover, queries and documents have different semantic meanings and should be dealt with differently. To this end, we propose to regard each query or document as a vertex and construct the \emph{query homogeneous graph} and \emph{document homogeneous graph} respectively, to model dependencies in user behaviors. Figure~\ref{fig:graph construction} shows an example of these two constructed homogeneous graphs. The query homogeneous graph consists of two kinds of edges:

\begin{figure}[t]
	\centering
	\includegraphics[width=0.45\textwidth]{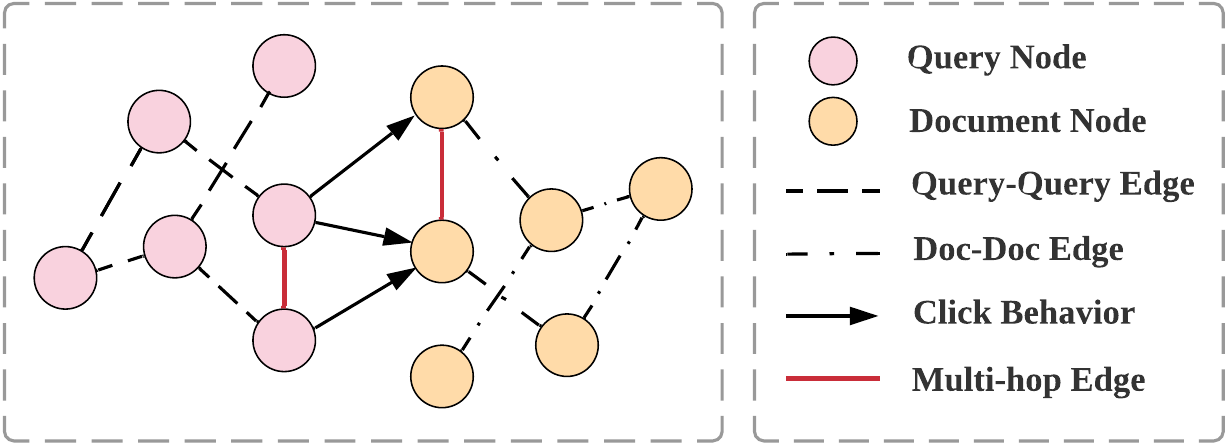}
    \caption{Graph construction methods in GraphCM.}
    \vspace{-10pt}
    \label{fig:graph construction}
\end{figure}

\begin{figure}[t]
	\centering
	\includegraphics[width=0.475\textwidth]{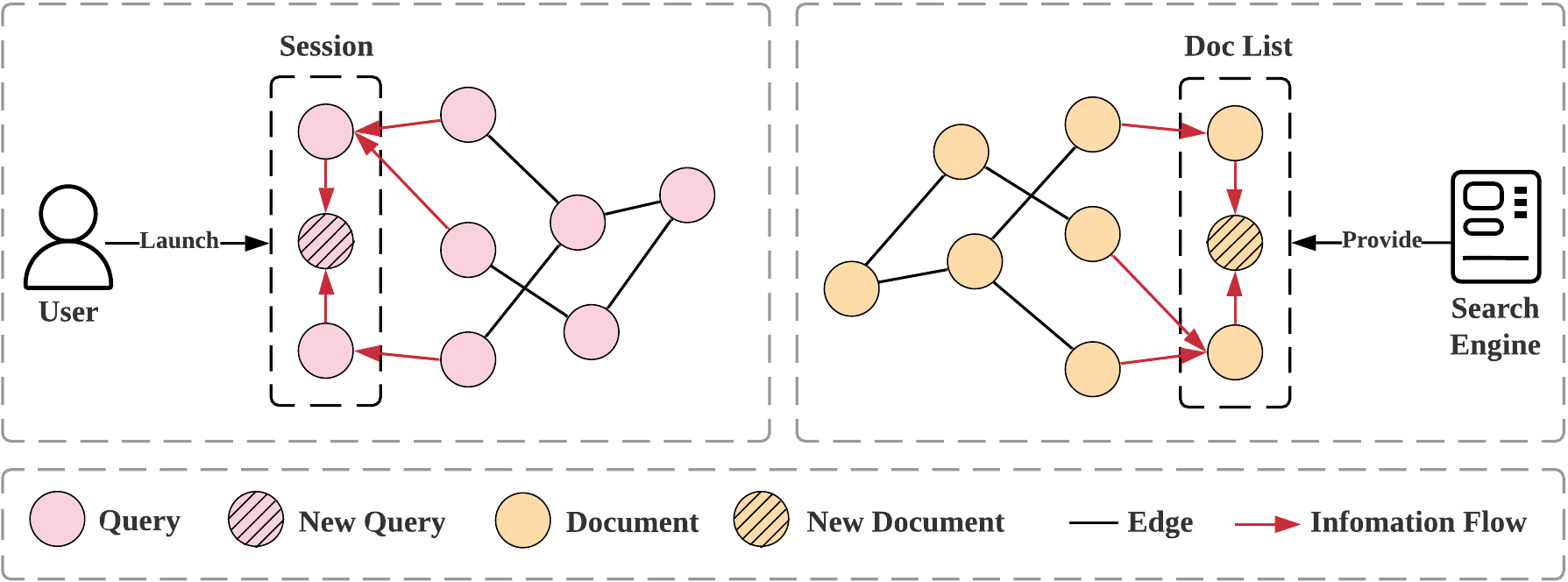}
    \caption{Graph solution for the cold-start problem.}
    \vspace{-15pt}
    \label{fig:cold start illustration}
\end{figure}

%Many existing studies have shown that incorporating the embeddings learned from the graph structure improves the system performance~\cite{jiang2018rin}. We regard each query or document as a vertex and construct the \emph{query homogeneous graph} and \emph{document homogeneous graph} respectively. Figure~\ref{fig:graph construction} shows an example of these two constructed homogeneous graphs. The query homogeneous graph consists of two kinds of edges:

\begin{itemize}[leftmargin=10pt]
    \item \textbf{Query Multi-hop Edge}. The edges between queries that click the same document, which denotes the potentially collaborative information between queries.
    \item \textbf{Query-Query Edge}. The edges between each pair of two consecutive queries in the same session, which denotes the reformulation relationships among queries issued by the user.
\end{itemize}
It is worth noting that the term \emph{reformulation} is different from the traditional IR task \emph{query reformulation}. Here, \emph{reformulation} means that the consecutive queries in the same session issued by a user should share similar intents towards similar topics. Likewise, the document homogeneous graph also contains two kinds of edges:

\begin{itemize}[leftmargin=10pt]
    \item \textbf{Doc Multi-hop Edge}. The edges between documents that are clicked by the same query, which denotes the potentially collaborative information between documents.
    \item \textbf{Doc-Doc Edge}. The edges between each pair of two consecutive documents in a ranked list, which denotes the similarity relationships among documents.
\end{itemize}

The graph construction methods above can fully exploit the intra-session and inter-session information (i.e., collaborative, reformulation and similarity information), which helps address the sparsity problem. As for the cold-start problem, a brand-new query or document vertex cannot have multi-hop edges since it has no click interactions available in the training set. As shown in Figure~\ref{fig:cold start illustration}, rather than make random guess as in previous click models (no previous click interactions is available), we can aggregate the reformulation or similarity information from the consecutive queries or documents to assist user click prediction, which mitigates the cold-start problem.

% As illustrated in Figure~\ref{fig:intra and inter session}, by the concept of \emph{collaborative filtering}, if queries across different sessions click the same document, or documents across different sessions are clicked by the same query, they should provide useful cross-session information, which is encoded by multi-hop edges, to cope with the sparsity problem and improve the system performance. Besides, previous works~\cite{qu2019end,wang2018ripplenet} have shown that applying graph neural networks can help tackle the cold-start problems by aggregating information from neighbors in graph structures.

\subsection{Embedding Layer}

GraphCM takes query $q$, document $d$, click variable $c$, vertical type $v$ and ranked position $p$ as inputs. Before the main process of the model, these original ID features are transformed into a high-dimensional sparse features via one-hot encoding. Then we apply embedding layers on the one-hot vectors to map them to low-dimensional dense embedding vectors:
\begin{equation}
\begin{aligned}
    \mathbf{v}_{q}=\mathbf{Emb}_{\mathbf{q}}\left(q\right),& 
    \mathbf{v}_{d}=\mathbf{Emb}_{\mathbf{d}}\left(d\right), \\
    \mathbf{v}_{c}=\mathbf{Emb}_{\mathbf{c}}\left(c\right),
    \mathbf{v}_{v}=\mathbf{Em}&\mathbf{b}_{\mathbf{v}}\left(v\right),
    \mathbf{v}_{p}=\mathbf{Emb}_{\mathbf{p}}\left(p\right),
\end{aligned}
\end{equation}
where $\mathbf{Emb}_{\mathbf{*}} \in \mathcal{R}^{N_{*} \times l_{*}},\; * \in \{\mathbf{q},\mathbf{d},\mathbf{c},\mathbf{v},\mathbf{p}\}$. 
$N_*$ and $l_*$ denote the input feature size and embedding size. For the ease of presentation, we omit the subscripts of embeddings when there is no ambiguity.

\subsection{Attractiveness Estimator}

The attractiveness estimator aims to estimate the attractiveness of each document $d_{i,j}$ to the user who issues the query $q_i$. For the pre-constructed homogeneous graphs, we propose a query encoder and a document encoder respectively to extract intra-session and inter-session information, and generate corresponding query and document context representations. Moreover, we propose a neighbor interaction module to further exploit the high-order neighbor information. Next, we will introduce these three components respectively (i.e., query encoder, document encoder and neighbor interaction module).

% The attractiveness estimator contains three components: a query encoder, a document encoder, and a neighbor interaction module. We will introduce these three components respectively.

\subsubsection{Query Encoder}
\label{sec:query encoder}

The connectivity in the query homogeneous graph mentioned in Section~\ref{sec:graph construction} provides collaborative and reformulation information among queries. Therefore, we first apply \emph{graph attention network}~\cite{Petar2018graph} (GAT) to aggregate neighbor information, and then generate query context representations via a Gated Recurrent Unit (GRU)~\cite{chung2014empirical}.
% \Jianghao{I think a newline is needed here. Same at the document encoder.}

After obtaining query vectors from the embedding layer, we apply GAT based on the query homogeneous graph to model structural dependencies among queries. Since the constructed graph is very large for real-world datasets, it is time consuming and not tractable to directly perform GAT. Therefore, for each query node $q_{i}$, we sample a fixed number (e.g., K) of neighbors to form its neighbor set $\mathcal{N}_{q_i}$. It is worth noting that, we add self-loop for every node to ensure that $q_i$ itself can be included in its neighbor set $\mathcal{N}_{q_i}$. Then, we aggregate the information of these neighbors to generate a new embedding $\mathbf{v}_{q_i}^{\prime}$ for query $q_i$ via a shared asymmetric attention mechanism:
\begin{equation}
\begin{aligned}
    \alpha_{i,j}=\operatorname{Softmax}_{j}&(\operatorname{att}_{\text{query}}(\mathbf{v}_{q_i},\mathbf{v}_{q_j})),\; q_j \in \mathcal{N}_{q_i}, \\
    \mathbf{v}_{q_i}^{\prime}=&\sum\nolimits_{q_{j} \in \mathcal{N}_{q_i}}\alpha_{i,j}\mathbf{v}_{q_j}.\\
\end{aligned}
\label{eq: query encoder attention}
\end{equation}

Here $\operatorname{att}_{\text{query}}$ denotes a linear layer with a $\operatorname{LeakyReLU}$ activation function, which performs the asymmetric node-level attention. $\operatorname{att}_{\text{query}}$ is shared for all the nodes in the query homogeneous graph to capture similar connection patterns. Moreover, similar to~\cite{vaswani2017attention}, we extend the proposed attention mechanism to \emph{multi-head attention} to stabilize the learning process of GAT. Specifically, $\mathcal{B}$ independent asymmetric node-level attention mechanisms execute the transformation of Equation~\ref{eq: query encoder attention}, and then their features are concatenated, resulting in the following output feature representations:
\begin{equation}
    \mathbf{v}_{q_i}^{\prime} = \mathop{\parallel}\limits_{b=1}^{\mathcal{B}}\sigma\left(\sum\nolimits_{q_{j} \in \mathcal{N}_{q_i}}\alpha_{i,j}^{b}\mathbf{v}_{q_j}
    \right),
\end{equation}
where $\mathop{\parallel}$ denotes vector concatenation, $\alpha^{b}_{i,j}$ is normalized attention coefficients computed by the $b$-th attention mechanism $\operatorname{att}_{\text{query}}^{b}$, $\sigma$ is the LeakyReLU function. Following~\cite{berg2017graph}, to further exploit multi-head attention mechanism, we propose another way for output feature aggregations. Instead of simple concatenation, we employ \emph{averaging aggregation}, and delay applying final non-linearity function on each attention mechanism until averaging is performed:
\begin{equation}
    \mathbf{v}_{q_i}^{\prime} = \sigma\Big(\frac{1}{\mathcal{B}}\sum_{b=1}^{\mathcal{B}}\sum_{q_{j} \in \mathcal{N}_{q_i}}\alpha_{i,j}^{b}\mathbf{v}_{q_j}\Big)\,.
\end{equation}
The number of heads and the selection on these two multi-head attention aggregation methods are hyperparameters in GraphCM, and should be fine-tuned according to different tasks. After the GAT process, we encode the sequence of query embeddings $\{\mathbf{v}_{q_1}^{\prime},...,\mathbf{v}_{q_n}^{\prime}\}$ through a standard Gated Recurrent Unit (GRU) to generate the query context representations:
\begin{equation}
\mathbf{h}_{i}^{q}=\operatorname{GRU}_{\text{query}}\left(\mathbf{v}_{q_1}^{\prime},\mathbf{v}_{q_2}^{\prime},...,\mathbf{v}_{q_i}^{\prime}\right),\; i=1,...,n.
\end{equation}
\subsubsection{Document Encoder}
\label{sec:doc encoder}
The document homogeneous graph mentioned in Section~\ref{sec:graph construction}, on the other hand, preserves collaborative and similarity information among documents. Similar to the query encoder, to extract useful clues for user behaviors from documents, we first perform neighbor aggregation via GAT, and then apply GRU to generate document context representations.
% Similar to the query encoder, the document encoder is designed for extracting useful clues for user behaviors from documents. 

The document encoder takes document ID $d$, vertical type $v$, previous click variable $c$ and ranked position $p$ as inputs and generates document context representations. Likewise, we apply the document homogeneous GAT on the document vectors after embedding layers. Neighbor sampling is also performed to form the neighbor sets $\mathcal{N}_{d_{i,j}}$ for $d_{i,j}$, which denotes the j-th document in the i-th query in the current session.
\begin{equation}
\begin{aligned}
    \beta_{i,j,k}=\operatorname{Softmax}_{k}&(\operatorname{att}_{\text{doc}}(\mathbf{v}_{d_{i,j}},\mathbf{v}_{d_k})),\; d_k \in \mathcal{N}_{d_{i,j}}, \\
    \mathbf{v}_{d_{i,j}}^{\prime}=&\sum\nolimits_{d_{k} \in \mathcal{N}_{d_{i,j}}}\beta_{i,j,k}\mathbf{v}_{d_k},\\
\end{aligned}
\end{equation}
where $\operatorname{att}_{\text{doc}}$ is an asymmetric node-level linear attention and is shared for all the nodes in the document homogeneous graph. We also apply a multi-head attention mechanism with two different feature aggregation methods to stabilize the learning process of the graph attention network:
\begin{equation}
\begin{aligned}
    &\mathbf{v}_{d_{i,j}}^{\prime} = \mathop{\parallel}\limits_{b=1}^{\mathcal{B}}\sigma\Big(\sum\nolimits_{d_{k} \in \mathcal{N}_{d_{i,j}}}\beta_{i,j,k}^{b}\mathbf{v}_{d_k}\Big),\; or \\
    &\mathbf{v}_{d_{i,j}}^{\prime} = \sigma\Big(\frac{1}{\mathcal{B}}\sum_{b=1}^{\mathcal{B}}\sum_{d_{k} \in \mathcal{N}_{d_{i,j}}}\beta_{i,j,k}^{b}\mathbf{v}_{d_k}\Big).
\end{aligned}
\end{equation}

After obtaining the document representations from GAT, we concatenate all input embeddings and generate document context representations through a GRU:
\begin{equation}
\begin{aligned}
\mathbf{x}_{i,j}&=[\mathbf{v}_{d_{i,j}}^{\prime} \oplus \mathbf{v}_{v_{i,j}} \oplus \mathbf{v}_{c_{i,j}} \oplus \mathbf{v}_{p_{i,j}}], \\
\mathbf{h}_{i,j}^{d}&=\operatorname{GRU}_{\text{doc}}\left(\mathbf{x}_{1,1},\mathbf{x}_{1,2},...,\mathbf{x}_{i,j}\right).
\end{aligned}
\end{equation}

Here $\oplus$ denotes the vector concatenation operation, which shares the same meaning with the symbol $\parallel$.

\subsubsection{Neighbor Interaction Module}

It is essential to take interactions between queries and documents into account in the field of click models. However, the previous two components (i.e., query encoder and document encoder) only model the context information of queries and documents separately.  We should further consider interactions between queries and documents. Moreover, instead of only considering the interaction between the current query $q_i$ and document $d_{i,j}$, we propose a neighbor interaction method to explicitly incorporate the document's high-order neighbor information, which further enrich the local graph structural information and alleviate the data sparsity problem. Neighbors of the query are not considered since the query homogeneous graph is more sparse than the document homogeneous graph.
Therefore, for the current query $q_i$ and document $d_{i,j}$, we first apply neighbor sampling on homogeneous GATs to generate the document's neighbor set $\mathcal{N}_{d_{i,j}}$ and adjusted embeddings $\mathbf{v}_q^{\prime},\mathbf{v}_d^{\prime}$, then we perform an attention-based feature interaction layer between the current query $q_{i}$ and neighbor documents $d_k$ in the neighbor set $\mathcal{N}_{d_{i,j}}$:
\begin{equation}
\begin{aligned}
&\mathbf{x}_{i,k}=\mathbf{v}_{q_{i}}^{\prime} \odot \mathbf{v}_{d_{k}}^{\prime},\; d_k \in \mathcal{N}_{d_{i,j}}, \\
&\gamma_{k}=\operatorname{Softmax}_{k}(\operatorname{att}_{\text{inter}}(\mathbf{x}_{i,k})),\; d_k \in \mathcal{N}_{d_{i,j}}, \\
&\mathbf{h}_{i,j}^{i}=\sum\nolimits_{d_k \in \mathcal{N}_{d_{i,j}}}\gamma_{k}\mathbf{x}_{i,k},\; d_k \in \mathcal{N}_{d_{i,j}},\\
\end{aligned}
\end{equation}
where $\odot$ denotes an element-wise product for vectors, and $\operatorname{att}_{\text{inter}}$ is a shared asymmetric linear attention layer with a LeakyReLU activation function. It is worth noting that we do not perform the multi-head attention mechanism in the neighbor interaction module. Previous works~\cite{qu2019end} and our empirical experiments have shown that the multi-head attention mechanism is not much of a help for improving the performance, while contributing to the computational cost.

After generating the query context representation $\mathbf{h}_{i}^{q}$, the document context representation $\mathbf{h}_{i,j}^{d}$ and the neighbor interaction representation $\mathbf{h}_{i,j}^{i}$, we concatenate them together and put them through a two-layer Multi-Layer Perceptron (MLP) to output the estimated attractiveness score $\mathcal{A}_{i,j}$ for the current query $q_i$ and document $d_{i,j}$:
\begin{equation}
\begin{aligned}
\mathbf{x}_{i,j}=[\mathbf{h}_{i}^{q}& \oplus \mathbf{h}_{i,j}^{d} \oplus \mathbf{h}_{i,j}^{i}], \\
\mathcal{A}_{i,j}=&\operatorname{MLP}(\mathbf{x}_{i,j}).
\end{aligned}
\end{equation}

Here, the activation functions for the first and the second layer in MLP are both LeakyReLU functions.

\subsection{Examination Predictor}

While the attractiveness estimator measures the attractiveness scores of each document to the user, the examination predictor aims to predict whether the user will continue to examine the document $d_{i,j}$ based on her session context. Following~\cite{chen2020context}, we assume that a user's examination action is only affected by her actions on previous documents --- the examination probability is not affected by the content of the document (users read the content only when the examination behavior happens). Therefore, for the current document $d_{n,m}$, we apply a session-level GRU to encode the information of ranked position $p$, vertical type $v$ and previous click variables $c$ within the same session. Finally, a linear layer followed by a Sigmoid function is performed on the encoded hidden states to output the examination probability $\mathcal{E}$:
\begin{equation}
\begin{aligned}
\mathbf{x}_{i,j}&=[\mathbf{v}_{p_{i,j}} \oplus \mathbf{v}_{v_{i,j}} \oplus \mathbf{v}_{c_{i,j}}],\\ 
\mathbf{h}_{i,j}^{e}&=\operatorname{GRU}_{exam}\left(\mathbf{x}_{1,1}, \mathbf{x}_{1,2},...,\mathbf{x}_{i,j}\right),\\
\mathcal{E}_{i,j} &= \operatorname{Sigmoid}\left(\operatorname{Linear}\left(\mathbf{h}_{i,j}^{e}\right)\right).
\end{aligned}
\end{equation}
\subsection{Click Predictor}

The click predictor module combines the examination probability $\mathcal{E}$ and attractiveness score $\mathcal{A}$, and outputs the predicted click probability. We implement four different combination functions, which are shown in Table~\ref{tab:combine func}.

\begin{table}[h]
    \centering
    \caption{Combination functions. E.H. is short for examination hypothesis. $\alpha$ and $\beta$ are learnable parameters.}
    \label{tab:combine func}
	\scalebox{0.9}{
	\renewcommand\arraystretch{1.2}
    \begin{tabular}{c c c}
    \toprule
    Function & Formula & Support E.H.? \\
    \midrule
    \emph{mul} & $c=\mathcal{E}\times \mathcal{A}$ & Yes \\
    \hline
    \emph{expmul} & $c=\mathcal{E}^{\alpha}\times \mathcal{A}^{\beta}$ & Yes \\
    \hline
    \emph{linear} & $c=\alpha \mathcal{E}+\beta \mathcal{A}$ & No \\
    \hline
    \emph{nonlinear} & $c=\operatorname{MLP}(\mathcal{E},\mathcal{A})$ & No \\
    \bottomrule
    \end{tabular}
    }
    \vspace{-10pt}
\end{table}

The \emph{mul} function simply follows the examination hypothesis and directly multiply the examination probability $\mathcal{E}$ with attractiveness score $\mathcal{A}$. The \emph{expmul} function preserves the examination hypothesis and increases the model capacity by adding learnable parameters. The \emph{linear} and \emph{nonlinear} functions further explore the relation between $\mathcal{E}$ and $\mathcal{A}$ beyond the examination hypothesis.

After click prediction, we adopt a binary cross-entropy loss to ensure an end-to-end training of GraphCM. The objective function to be minimized during training is:
\begin{equation}
\small{
    \mathcal{L}(\theta)=\mathcal{L}_{c}(\theta)+\lambda \| \theta \|^2
}
\end{equation}
\begin{equation}
\small{
    \mathcal{L}_c(\theta)=-\frac{1}{N}\sum_{i,j}C_{i,j}\log \mathcal{P}_{i,j}+(1-C_{i,j})\log(1-\mathcal{P}_{i,j})
}
\end{equation}
where $\theta$ denotes trainable parameters in GraphCM, $\lambda$ denotes the hyperparameter for regularization, $N$ denotes the number of training batches, $C_{i,r}$ and $\mathcal{P}_{i,j}$ denote the real click signal and the predicted click probability.

\section{Experiment}

In this section, we conduct extensive experiments to answer the following questions:

\begin{itemize}[leftmargin=25pt]
    \item[\textbf{RQ1}] Does GraphCM mitigate the data sparsity problem and achieve the best performance compared with baseline models?
    \item[\textbf{RQ2}] Can GraphCM tackle the cold-start problem? How does GraphCM perform if it meets a brand-new query or document during the test phase?
    \item[\textbf{RQ3}] Which combination function performs best in integrating the attractiveness scores and examination probabilities?
    \item[\textbf{RQ4}] What is the influence of different components in GraphCM?
\end{itemize}

\subsection{Experimental Setup}

\subsubsection{Dataset}

We choose three real-world public session 
datasets in web search collected by different search engine platforms: Yandex\footnote{\url{https://www.kaggle.com/c/yandex-personalized-web-search-challenge}}, TREC2014\footnote{\url{https://trec.nist.gov/data/session2014.html}} and TianGong-ST\footnote{\url{http://www.thuir.cn/tiangong-st/}}~\cite{chen2019tiangong}. The statistics of the datasets can be found in Table~\ref{tab:dataset statistics}. Due to the memory limitation, we downsample the size of Yandex dataset. All datasets are divided into training, validation, and test sets with proportion 8:1:1. Besides, we would like to make two statements about the datasets before we elaborate on the details of the experiments:

\begin{itemize}[leftmargin=18pt]
    \item[(1)] TianGong-ST is the only dataset that provides the vertical type information for each document in the search logs. For other datasets, we simply assume that all documents are presented in the same vertical type, and assign the same vector embedding for all vertical types.
    % \item[(2)] Yahoo dataset does not provide session-level information. That is, the session $\mathcal{S}$ issued by the user only contains one query. As shown in Table~\ref{tab:dataset statistics}, the number of sessions is equal to the number of queries. This brings critical issues for GraphCM to tackle the cold-start problem, which will be discussed in Section~\ref{sec:cold-start}
    \item[(2)] TianGong-ST is the only dataset that provides relevance labels for query-document pairs. Therefore, we perform the relevance estimation task only on the TianGong-ST dataset, and perform the click prediction task on all the three datasets.
\end{itemize}

\begin{table}[h]
    \centering
    \vspace{-6pt}
    \caption{The dataset statistics}
    \vspace{-3pt}
    \label{tab:dataset statistics}
% 	\resizebox{0.45\textwidth}{!}{
% 	\renewcommand\arraystretch{1.05}
    \begin{tabular}{c c c c}
    \toprule
    Dataset & Session & Query & Search Engine \\
    \midrule
    Yandex & 200,000 & 376,965 & Yandex \\
    % \hline
    TREC2014 & 1,257 & 5,443 & Indri \\
    % \hline
    TianGong-ST & 147,155 & 356,252 & Sogou \\
    \bottomrule
    \end{tabular}
    \vspace{-6pt}
    % }
\end{table}

\begin{table*}[h]
	\centering
	\caption{Overall performance of each click model. We only perform relevance estimation task on TianGong-ST, since it is the only dataset that provides human-annotated relevance labels. The best results are given in bold, while the second best values are underlined. $*$ indicates statistically significant improvement (measured by t-test) with p-value $<$ 0.001 over all baselines. \emph{Note}: LL, the higher, the better; PPL, the lower, the better; NDCG, the higher, the better.}
	\label{tab:overall performance}
	\resizebox{0.95\textwidth}{!}{
	\renewcommand\arraystretch{1.25}
	\begin{tabular}{c|c c|c c|c c c c c c}
		\toprule
		\hline
		\multicolumn{1}{c|}{\multirow{2}{*}{Model}} & \multicolumn{2}{c|}{\textbf{Yandex}} &  \multicolumn{2}{c|}{\textbf{TREC2014}} &  \multicolumn{6}{c}{\textbf{TianGong-ST}} \\
		\cline{2-11}
		\multicolumn{1}{c|}{} & LL & PPL & LL & PPL & LL & PPL & NDCG@1 & NDCG@3 & NDCG@5 & NDCG@10 \\
		\cline{1-11}
		\multicolumn{1}{c|}{CCM} & -0.4171 & 1.3436 & -0.5445 & 1.3267 & -0.2057  & 1.2238 & 0.6596 & 0.6931 & 0.7175 & 0.8453 \\
		\multicolumn{1}{c|}{DCM} & -0.4183 & 1.3226 & -0.5485 & 1.2998 & -0.2109  & 1.2231 & 0.6861 & 0.6818 & 0.7115 & 0.8441 \\
		\multicolumn{1}{c|}{DBN} & -0.4157 & 1.3397 & -0.5451 & 1.3216 & -0.2052  & 1.2292 & 0.6883 & 0.6997 & 0.7261 & 0.8501 \\
		\multicolumn{1}{c|}{SDBN} & -0.4174 & 1.3399 & -0.5491 & 1.3218 & -0.2094  & 1.2293 & 0.6814 & 0.6777 & 0.7104 & 0.8421 \\
		\multicolumn{1}{c|}{PBM} & -0.2429 & 1.2961 & -0.1678 & 1.1899 & -0.1825  & 1.2183 & 0.6594 & 0.6422 & 0.6703 & 0.8243 \\
		\multicolumn{1}{c|}{UBM} & -0.2275 & 1.2662 & -0.1568 & 1.1901 & -0.1751  & 1.2179 & 0.6362 & 0.6354 & 0.6671 & 0.8208 \\
		\hline
 		\multicolumn{1}{c|}{NCM} & -0.2204 & 1.2666 & -0.1549 & 1.1757 & -0.1718  & 1.2055 & 0.7081 & 0.7066 & 0.7368 & 0.8621 \\
 		\multicolumn{1}{c|}{CACM} & \underline{-0.2199} & \underline{1.2662} & \underline{-0.1541} & \underline{1.1738} & \underline{-0.1707}  & \underline{1.2027} & \underline{0.7347} & \underline{0.7163} & \underline{0.7405} & \underline{0.8667} \\
 		\multicolumn{1}{c|}{GraphCM} & $\textbf{-0.2192}^{*}$ & $\textbf{1.2652}^{*}$ & $\textbf{-0.1529}^{*}$ & $\textbf{1.1721}^{*}$ & $\textbf{-0.1629}^{*}$ & $\textbf{1.1938}^{*}$ & $\textbf{0.7388}^{*}$ & $\textbf{0.7189}^{*}$ & $\textbf{0.7466}^{*}$ & $\textbf{0.8671}^{*}$ \\
 		\hline
 		\bottomrule
	\end{tabular}
	}
\end{table*}

\begin{table}[t]
	\centering
	\caption{Data sparsity statistics and LL/PPL improvements. The sparsity is calculated as \# missing query-document interactions / \# possible query-document pairs. The improvements are absolute LL/PPL gains of GraphCM compared with the best baseline.}
	\label{tab:sparsity and improvement}
	\resizebox{0.45\textwidth}{!}{
	\renewcommand\arraystretch{1.1}
	\begin{tabular}{c c c c}
		\toprule
		Dataset & Sparsity & LL Imprv. & PPL Imprv. \\
		\midrule
		Yandex & 99.8895\% & 0.32\% & 0.08\% \\
		TREC2014 & 99.9189\% & 0.78\% & 0.14\% \\
		TianGong-ST & 99.9969\% & 4.57\% & 0.74\% \\
 		\bottomrule
	\end{tabular}
	}
    % \vspace{-14pt}
\end{table}

\subsubsection{Baselines}

The existing click models can be categorized into two classes: PGM-based and NN-based methods. In this paper, We consider CCM~\cite{Guo2009CCM}, DCM~\cite{Guo2009DCM}, DBN~\cite{Chapelle2009DBN}, SDBN~\cite{chuklin2015click}, PBM~\cite{craswell2008experimental} and UBM~\cite{dupret2008user} as representative PGM-based click models, of which open-source implementations are available\footnote{\url{https://github.com/markovi/PyClick}}. For NN-based click models, we consider NCM~\cite{borisov2016neural} and CACM~\cite{chen2020context} as baselines.

\subsubsection{Evaluation Metrics}

We compare GraphCM with baseline models based on the following two tasks: \emph{click prediction} and \emph{relevance estimation}, which are both common tasks for click models~\cite{chuklin2015click}. For click prediction task, we report the log-likelihood (LL) and perplexity (PPL)~\cite{borisov2016neural} of each model. The definitions of the log-likelihood and click perplexity at the rank $r$ are as follows:
\begin{equation}
\small{
    LL = \frac{1}{MN}\sum_{i=1}^{N}\sum_{r=1}^{M}C_{i,r}\log \mathcal{P}_{i,r}+(1-C_{i,j})\log(1-\mathcal{P}_{i,r})
    }
\end{equation} 
\begin{equation}
\small{
    PPL@r=2^{-\frac{1}{N}\sum_{i=1}^{N}c_{i,r}\log\mathcal{P}_{i,r}+(1-C_{i,r})\log(1-\mathcal{P}_{i,r})}
}
\end{equation}
where subscript $r$ is the ranked position, $N$ is the total number of queries, and $M$ is the number of documents in each query. $C_{i,r}$ and $\mathcal{P}_{i,r}$ are the real click signal and the predicted click probability of the $r$-th document in the $i$-th query. We calculate the total perplexity by averaging perplexities over all the positions. Lower values of perplexity and higher values of log-likelihood correspond to better click prediction performance.

For the relevance estimation task, we use click models to rank the document list and compute the averaged Normalized Discounted Cumulative Gain (NDCG)~\cite{NDCG} according to the human-annotated relevance labels. We report NDCG scores at truncation level 1, 3, 5, and 10. Higher values of NDCG indicates better relevance estimation performance.

\subsubsection{Implementation Details}
We train GraphCM with a mini-batch size of 128 with Adam optimizer. The hidden sizes of all GRUs are 64. The embedding sizes of query $q$, document $d$, vertical type $v$, click variable $c$ and ranked position $p$ are 64, 64, 8, 4, 4 respectively. The initial learning rate is selected from $\{10^{-3},5\times10^{-4},10^{-4}\}$. To avoid overfitting, we choose the coefficient of L2 norm and dropout rate from $\{10^{-4}, 10^{-5}\}$ and $\{0.25, 0.5\}$. The number of neighbors to be sampled is tuned from $\{1, 2, 4, 8, 16, 32\}$.
Finally, we adopt the model at the iteration with the lowest validation PPL for evaluation in the test set. To ensure fair comparison, we also fine-tune all the baseline models to achieve their best performance. Our model implemented on PyTorch is available \footnote{\url{https://github.com/CHIANGEL/GraphCM}}.

\begin{table*}[ht]
	\centering
	\caption{LL (the higher, the better) and PPL (the lower, the better) performance of click models for the cold-start problems on different dataset. The best results are given in bold, while the second best values are underlined. $*$ indicates statistically significant improvement (measured by t-test) with p-value $<$ 0.001 over all baselines.}
	\label{tab:cold-start}
	\resizebox{0.95\textwidth}{!}{
	\renewcommand\arraystretch{1.3}
	\begin{tabular}{c|c|c c c c|c c c c|c c c c}
		\toprule
		\hline
		\multicolumn{1}{c|}{\multirow{2}{*}{Metric}} & \multicolumn{1}{c|}{\multirow{2}{*}{Model}} &  \multicolumn{4}{c|}{\textbf{Yandex}} &  \multicolumn{4}{c|}{\textbf{TREC2014}} &  \multicolumn{4}{c}{\textbf{TianGong-ST}}  \\
		\cline{3-14}
		\multicolumn{1}{c|}{} & \multicolumn{1}{c|}{} & Cold Q & Cold D & Cold QD & Warm QD & Cold Q & Cold D & Cold QD & Warm QD & Cold Q & Cold D & Cold QD & Warm QD\\
		\cline{1-14}
		\multicolumn{1}{c|}{\multirow{4}{*}{LL}} & \multicolumn{1}{c|}{UBM} & -0.2418 & -0.2397 & -0.2346 & -0.1755 & -0.3105 & -0.2021 & -0.1408 & -0.1432 & -0.2157 & -0.2011 & -0.2189 & -0.1696 \\
 		\multicolumn{1}{c|}{} & \multicolumn{1}{c|}{NCM} & -0.2386 & -0.1841 & -0.2343 & -0.1685 & -0.3069 & -0.1717 & -0.1398 & -0.1422 & -0.1987 & -0.2009 & -0.2091 & -0.1609 \\
 		\multicolumn{1}{c|}{} & \multicolumn{1}{c|}{CACM} & \underline{-0.2319} & \underline{-0.1798} & \underline{-0.2338} & \underline{-0.1626} & \underline{-0.3045} & \underline{-0.1711} & \underline{-0.1389} & \underline{-0.1421} & \underline{-0.1977} & \underline{-0.1979} & \underline{-0.2076} & \underline{-0.1586} \\
 		\multicolumn{1}{c|}{} & \multicolumn{1}{c|}{GraphCM} & $\textbf{-0.2316}^{*}$ & $\textbf{-0.1792}^{*}$ & $\textbf{-0.2324}^{*}$ & $\textbf{-0.1624}^{*}$ & $\textbf{-0.2954}^{*}$ & $\textbf{-0.1645}^{*}$ & $\textbf{-0.1379}^{*}$ & $\textbf{-0.1418}^{*}$ & $\textbf{-0.1948}^{*}$ & $\textbf{-0.1977}^{*}$ & $\textbf{-0.2074}^{*}$ & $\textbf{-0.1527}^{*}$ \\
 		\hline
		\multicolumn{1}{c|}{\multirow{4}{*}{PPL}} & \multicolumn{1}{c|}{UBM} & 1.3265 & 1.3174 & 1.3269 & 1.2242 & 1.4008 & 1.2234 & 1.1608 & 1.1688 & 1.2473 & 1.2689 & 1.2679 & 1.2082 \\
 		\multicolumn{1}{c|}{} & \multicolumn{1}{c|}{NCM} & 1.2931 & 1.2254 & 1.2847 & 1.2044 & 1.3967 & 1.2153 & 1.1597 & 1.1679 & 1.2363 & 1.2441 & 1.2539 & 1.1919 \\
 		\multicolumn{1}{c|}{} & \multicolumn{1}{c|}{CACM} & \underline{1.2829} & \underline{1.2185} & \underline{1.2839} & \underline{1.1957} & \underline{1.3564} & \underline{1.2106} & \underline{1.1581} & \underline{1.1677} & \underline{1.2355} & \underline{1.2392} & \underline{1.2513} & \underline{1.1885} \\
 		\multicolumn{1}{c|}{} & \multicolumn{1}{c|}{GraphCM} & $\textbf{1.2823}^{*}$ & $\textbf{1.2176}^{*}$ & $\textbf{1.2823}^{*}$ & $\textbf{1.1956}^{*}$ & $\textbf{1.2381}^{*}$ & $\textbf{1.2013}^{*}$ & $\textbf{1.1578}^{*}$ & $\textbf{1.1676}^{*}$ & $\textbf{1.2336}^{*}$ & $\textbf{1.2389}^{*}$ & $\textbf{1.2508}^{*}$ & $\textbf{1.1792}^{*}$ \\
 		\hline
 		\bottomrule
	\end{tabular}
	}
\end{table*}

\subsection{Performance Comparison (RQ1)}

We perform click prediction and relevance estimation tasks on each click model for performance comparison. It is worth noting that the relevance estimation task is only performed on TianGong-ST dataset, as it is the only dataset that provides human-annotated relevance labels. The results are presented in Table~\ref{tab:overall performance}, from which we can obtain the following observations:

\begin{itemize}[leftmargin=18pt]
    \item[(1)] All NN-based models show significant improvements over PGM-based models in click prediction and relevance estimation tasks. NN-based models adopt the distributed vector representation approach for representations of queries and documents, therefore can better capture user behavior patterns.
    \item[(2)] CACM achieves the best performance among all the baseline models, followed by NCM and other PGM-based click models. CACM utilizes complex model structures to take the intra-session information into consideration and thus better capture the user behavior patterns, which is consistent with the  results reported in~\cite{chen2020context}.
    \item[(3)] GraphCM significantly outperforms all the baseline models. Such improvement validates the effectiveness of applying graph neural networks and neighbor interaction techniques to make use of the inter-session and intra-session information, which enables GraphCM to capture more subtle patterns in user click behaviors.
\end{itemize}

Following~\cite{qu2019end}, to study the ability of GraphCM to tackle the data sparsity problem, we further investigate the data sparsity statistics and improvements of GraphCM compared to the best baseline model among different datasets. The results are shown in Table~\ref{tab:sparsity and improvement}. As the data sparsity ratios increase (Yandex: 99.8895\%; TREC2014: 99.9189\%; TianGong-ST: 99.9969\%;), the GraphCM gradually achieves better improvements for LL and PPL metrics. This means that GraphCM exploits intra-session and inter-session information from the pre-constructed homogeneous graphs via graph neural networks and neighbor interaction techniques, thus leading to better performance in mitigating the data sparsity problem.

\subsection{Cold-start Problem (RQ2)}
\label{sec:cold-start}

Similar to works in recommender systems, existing click models suffer from the cold-start problem. Namely, there are brand-new queries or documents during the test phase that never appear in the training set. To ensure the unambiguity, before we zoom into the details of experiments, we first define the following concepts. Suppose we have a session $\mathcal{S}$ in the test set, which consists of a sequence of queries $\mathcal{Q}=\{q_1,...,q_n\}$ and documents $\mathcal{D}=\{d_{1,1},...,d_{n,m}\}$, we propose the following definitions:

\begin{definition}
Define a query $q$ in the test set as a \textbf{cold query} if and only if it never appears in the training set.
\label{def:cold query}
\end{definition}

\begin{definition}
Define a query $q$ in the test set as a \textbf{warm query} if and only if it has appeared in the training set.
\label{def:warm query}
\end{definition}

\begin{definition}
Define the session $\mathcal{S}$ as a \textbf{cold-query session} if and only if there exists at least one \emph{cold query} $q_i \in \mathcal{Q}$.
\label{def:cold-query session}
\end{definition}

\begin{definition}
Define the session $\mathcal{S}$ as a \textbf{warm-query session} if and only if all the queries $q_i \in \mathcal{Q}$ are \emph{warm queries}.
\label{def:warm-query session}
\end{definition}

% \begin{definition}
% Suppose there are $N_c$ cold queries and $N_w$ warm queries in the full test set, the \textbf{cold-query ratio} of the dataset is defined as $CQR=N_c / (N_c+N_w)$.
% \label{def:cold-query ratio}
% \end{definition}

Note that all the definitions above focus on the \emph{query cold-start} problem (i.e., new queries in the test set). We can further propose the parallel concepts for the \emph{document cold-start} problem, which is omitted here due to page limitations. Based on the definitions above, as illustrated in Figure~\ref{fig:test set division}, we can split the full test set into four mutually exclusive session sets: (\romannumeral1) Cold Q set; (\romannumeral2) Cold D set; (\romannumeral3) Cold QD set; (\romannumeral4) Warm QD set. The four session sets have varying degrees on the cold-start problem. Then we measure the LL and PPL performance of GraphCM, CACM, NCM and UBM on these session sets respectively for different datasets (NDCG metric is omitted since it shows similar trends). The results are shown in Table~\ref{tab:cold-start}, from which we can obtain the following observations:

\begin{figure}[t]
	\centering
	\includegraphics[width=0.46\textwidth]{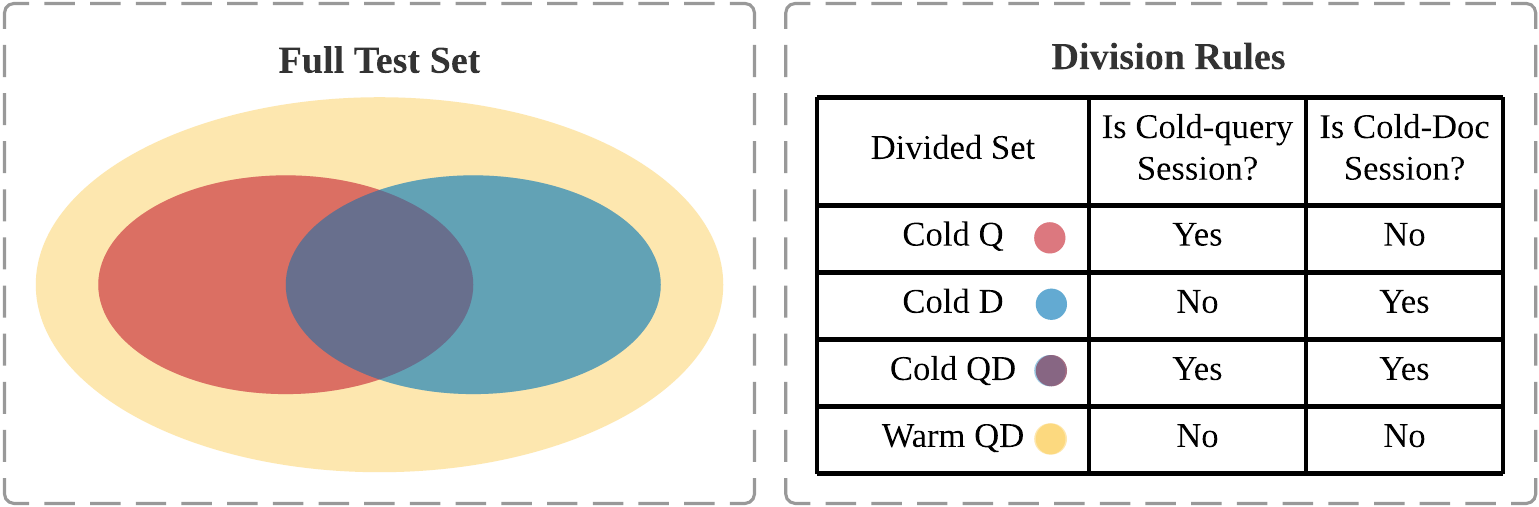}
    \caption{The full test set division for cold-start problems.}
    \vspace{-11pt}
    \label{fig:test set division}
\end{figure}

\begin{itemize}[leftmargin=18pt]
    \item[(1)] Generally, click models achieve much worse performance on the Cold Q set, Cold D set and Cold QD set compared to the performance on the Warm QD set, which validates the fact that click models are quite vulnerable to the cold-start problems. However, for TREC2014 dataset, we observe that the performance of click models on Cold QD set is better than it on Warm QD set. A possible reason is that Cold QD session dominates the validation set of TREC2014 dataset (up to 62.91\%). The model selection on such a validation set let the model fit the Cold QD data better.
    %Since the Cold QD session ratio in the validation set of TREC2014 dataset is up to 62.91\%, if we make model selection according to the validation performance, we may end up with a model that sacrifices the performance on Warm QD Set and obtains better performance on Cold QD set.
    \item[(2)] As a PGM-based model, UBM achieves the worst performance compared to NN-based methods on the four session sets, which suggests that the distributed vector representation approach adopted by NN-based methods is better than the traditional binary random variable representations in cold-start problems.
    \item[(3)] CACM achieves relatively better performance than NCM on the four session sets for three datasets. As CACM applies GRU units to encode the intra-session information during the test phase, it can somehow implicitly utilize the reformulation information between queries and similarity information between documents, which helps cope with the cold-start problems.
    \item[(4)] GraphCM outperforms all baselines on four session sets for three datasets. Different from CACM's implicit utilization of the reformulation and similarity information, GraphCM directly model these relations in the homogeneous graphs, and applies graph neural networks and neighbor interaction techniques to explicitly leverage the auxiliary information as guidance for user behavior modeling and click prediction. The results demonstrate that GraphCM can better tackle the cold start problems.
    \item[(5)] The full test set, as a union of four session sets, is more similar to real-world application scenarios where cold queries or documents are likely to show up among the warm ones. As presented in Table~\ref{tab:overall performance}, GraphCM achieves the best performance on the full test set, which implies the effectiveness of GraphCM to tackle the cold-start problem in real-world applications.
\end{itemize}

\begin{figure*}[ht]
	\centering
	\includegraphics[width=0.3\textwidth]{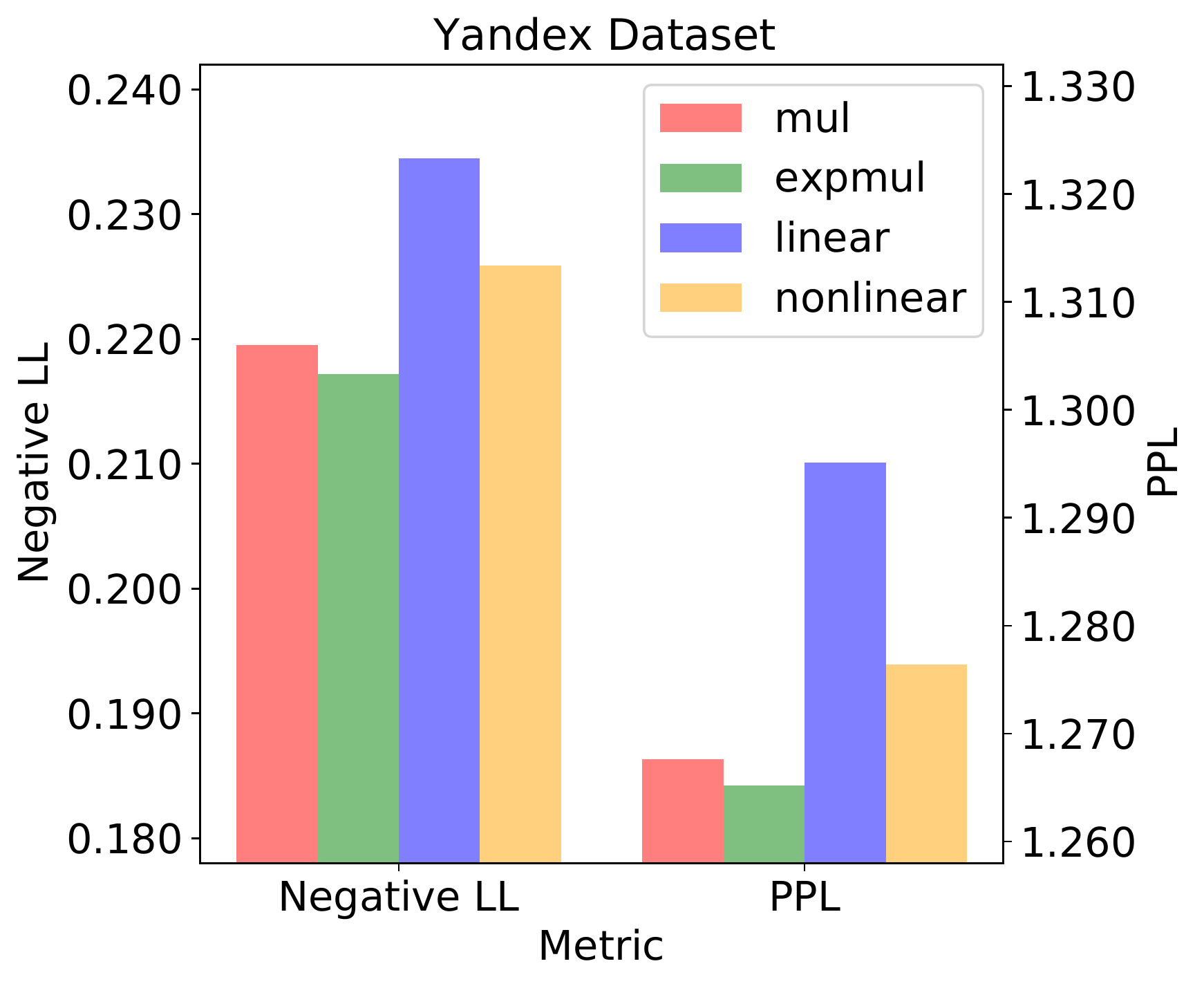}
	\includegraphics[width=0.3\textwidth]{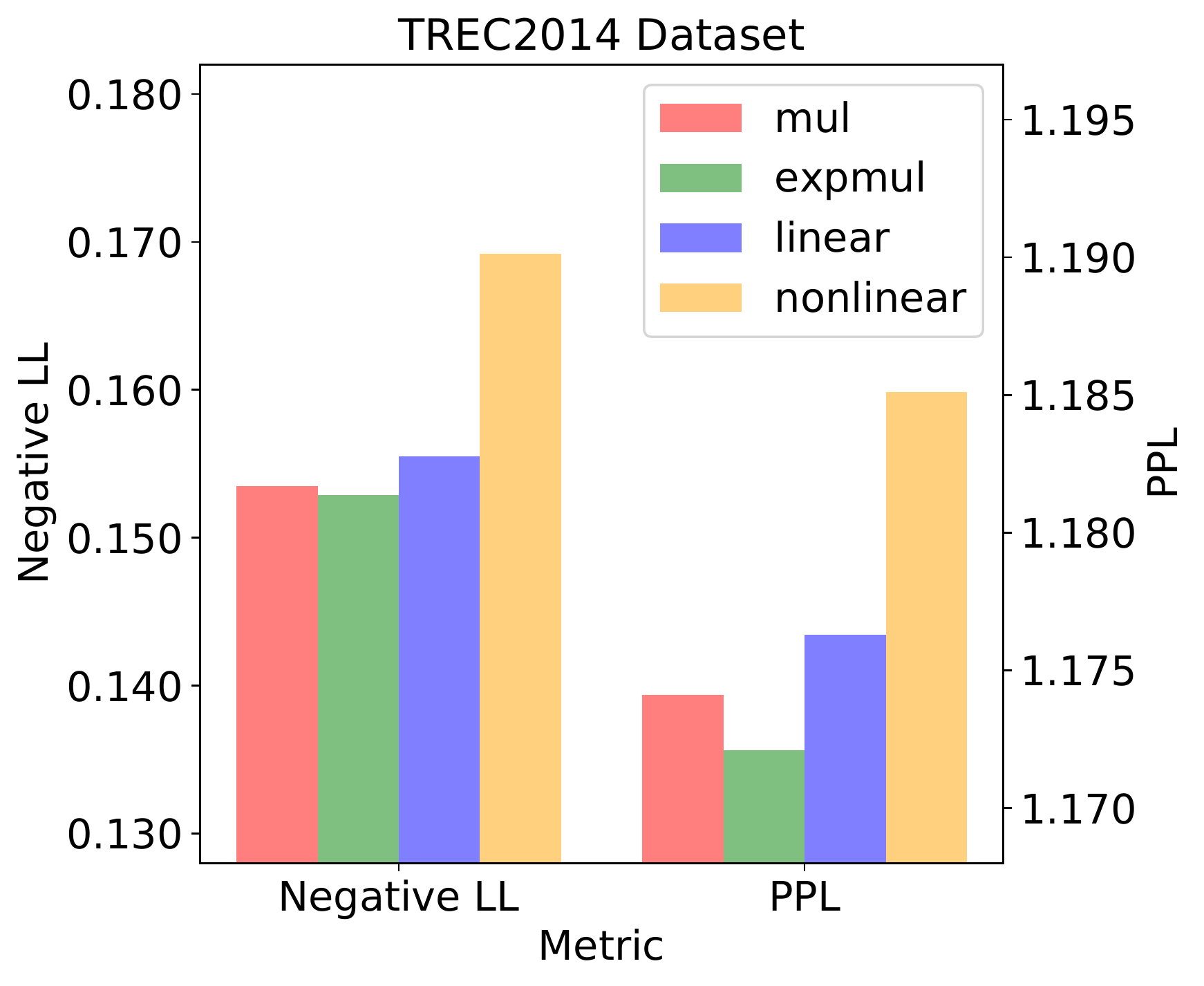}
	\includegraphics[width=0.3\textwidth]{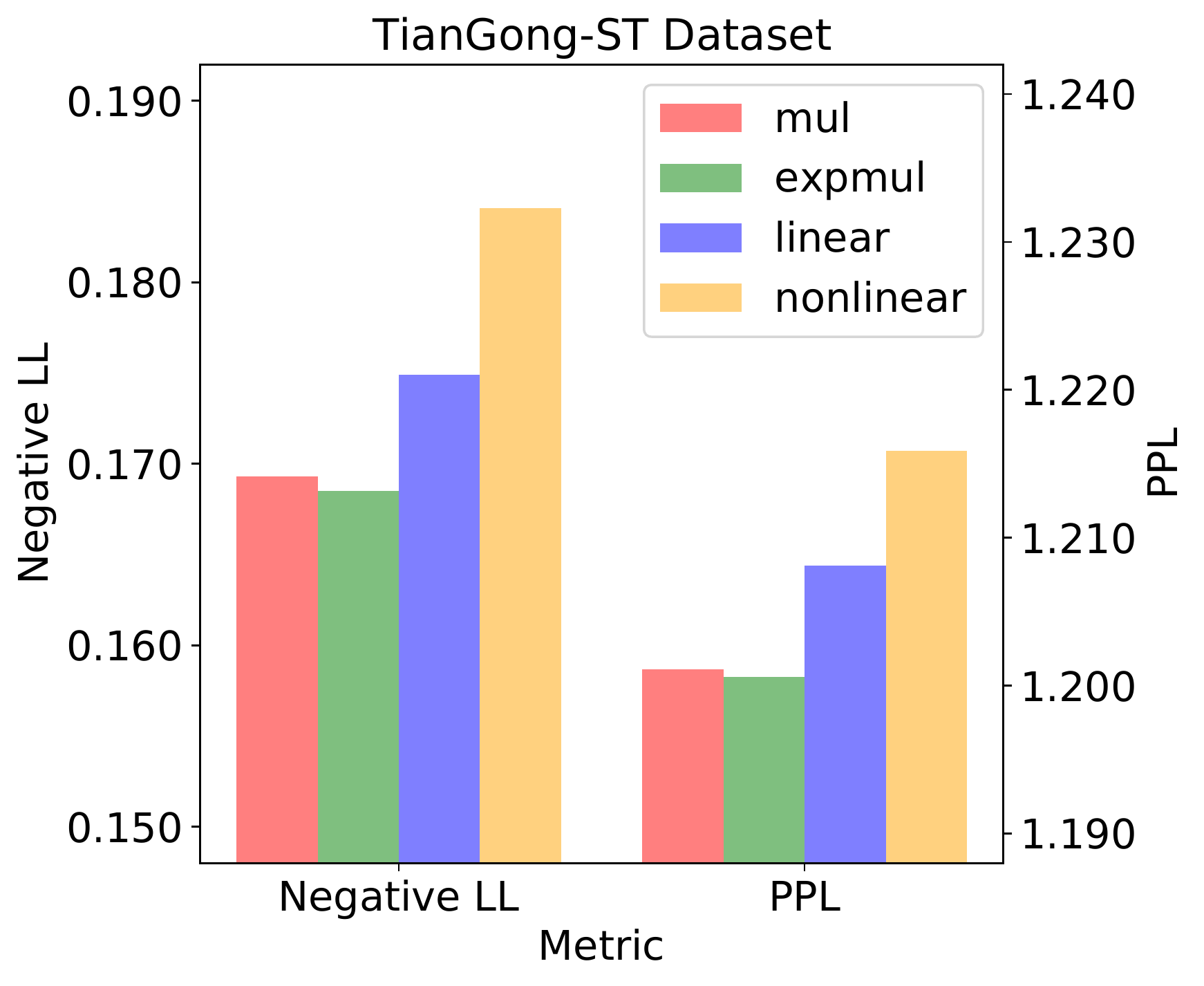}
	\vspace{-11pt}
	\caption{GraphCM's click prediction performance with different combination functions. \emph{Note}: Negative LL, the lower, the better; PPL, the lower, the better.}
	\vspace{-5pt}
	\label{fig:combine function}
\end{figure*}

\vspace{-2pt}
\subsection{Combination Function (RQ3)}

We study the effectiveness of different combination functions by comparing their click prediction performance, which is shown in Figure~\ref{fig:combine function}. From Figure~\ref{fig:combine function}, we can obtain the following observations:

\begin{itemize}[leftmargin=18pt]
    \item[(1)] GraphCMs with the \emph{mul} and \emph{expmul} functions outperform those with the \emph{linear} and \emph{nonlinear} functions. This indicates that combination functions that support the examination hypothesis can properly represent user behavior patterns and achieve better performance.
    \item[(2)] GraphCM with \emph{expmul} function achieves the best performance among all combination functions. Although the \emph{expmul} and \emph{mul} functions both support the examination hypothesis, the \emph{mul} function is only a special case of the \emph{expmul} function (i.e., $\alpha=1,\beta=1$). Therefore, the \emph{expmul} function can model user behaviors more flexibly with more learnable parameters.
    \item[(3)] In theory, the \emph{nonlinear} function, as a two-layer fully connected network, is able to cover the computational formulas of other combination functions and perform at least as good as other functions. However, the performance of \emph{nonlinear} function shown in Figure~\ref{fig:combine function} is much worse than other functions. We argue that the \emph{nonlinear} function overfits on the click signals and fails to learn the multiplication transformation that supports the examination hypothesis, resulting in poor performance.
\end{itemize}

% \begin{table}[t]
% 	\centering
% 	\caption{GraphCM's click prediction performance with different combination functions. Bold values are the best in each column, while the second best values are underlined. \emph{Note}: LL, the higher, the better; PPL, the lower, the better.}
% 	\label{tab:combine click performance}
% 	\resizebox{0.475\textwidth}{!}{
% 	\renewcommand\arraystretch{1.3}
% 	\begin{tabular}{c|c c|c c|c c}
% 	    \toprule
% 		\hline
% 		\multicolumn{1}{c|}{\multirow{2}{*}{Function}} &  \multicolumn{2}{c|}{\textbf{Yandex}} &  \multicolumn{2}{c|}{\textbf{TREC2014}} &  \multicolumn{2}{c}{\textbf{TianGong-ST}} \\
% 		\cline{2-7}
% 		\multicolumn{1}{c|}{} & LL & PPL & LL & PPL & LL & PPL \\
% 		\cline{1-7}
% 		\multicolumn{1}{c|}{\emph{mul}} & \underline{-0.2195} & \underline{1.2656} & \underline{-0.1535} & \underline{1.1741} & \underline{-0.1693}  & \underline{1.2011} \\
% 		\multicolumn{1}{c|}{\emph{expmul}} & \textbf{-0.2192} & \textbf{1.2652} & \textbf{-0.1529} & \textbf{1.1721} & \textbf{-0.1629}  & \textbf{1.1938} \\
% 		\multicolumn{1}{c|}{\emph{linear}} & -0.2545 & 1.3251 & -0.1555 & 1.1763 & -0.1749 & 1.2081 \\
% 		\multicolumn{1}{c|}{\emph{nonlinear}} & -0.2259 & 1.2764 & -0.1692 & 1.1911 & -0.1841 & 1.2259 \\
%  		\hline
%  		\bottomrule
% 	\end{tabular}
% 	}
% \end{table}

In function \emph{expmul} and \emph{linear}, the parameters $\alpha$ and $\beta$ are learnable, instead of hyperparameters that require manual assignments. To further investigate their learning mechanism, we check the values of the learnable parameters in functions \emph{expmul} and \emph{linear}. The results are shown in Table~\ref{tab:combine alpha beta}. We observe that $\alpha$ is always higher than $\beta$ for different combination functions and datasets. This indicates that GraphCM assigns higher weights to the attractiveness scores generated by the attractiveness estimator, since the structure of attractiveness estimator is more complex and is able to encode more important information for user behavior modeling.

\begin{table}[t]
	\centering
	\caption{The learnable parameters $\alpha$ and $\beta$ for combination functions $expmul$ and $linear$.}
	\label{tab:combine alpha beta}
	\resizebox{0.47\textwidth}{!}{
	\renewcommand\arraystretch{1.3}
	\begin{tabular}{c|c|c c|c c|c c}
	    \toprule
		\hline
		\multicolumn{1}{c|}{\multirow{2}{*}{Function}} &  \multicolumn{1}{c|}{\multirow{2}{*}{Formula}} &  \multicolumn{2}{c|}{\textbf{Yandex}} &  \multicolumn{2}{c|}{\textbf{TREC2014}} &  \multicolumn{2}{c}{\textbf{TianGong-ST}} \\
		\cline{3-8}
		\multicolumn{1}{c|}{} & \multicolumn{1}{c|}{} & $\alpha$ & $\beta$ & $\alpha$ & $\beta$ & $\alpha$ & $\beta$ \\
		\cline{1-8}
		\multicolumn{1}{c|}{\emph{expmul}} & $c=\mathcal{E}^{\alpha}\times \mathcal{A}^{\beta}$ & 1.035 & 0.991 & 1.031 & 1.010 & 0.952 & 0.796 \\
		\multicolumn{1}{c|}{\emph{linear}} & $c=\alpha \mathcal{E}+\beta \mathcal{A}$ & 0.468 & 0.462 & 0.463 & 0.437 & 0.491 & 0.401 \\
 		\hline
 		\bottomrule
	\end{tabular}
	}
% 	\vspace{-20pt}
\end{table}

\subsection{Ablation Study (RQ4)}

In order to investigate the contribution of each component to the final performance of GraphCM, we conduct several comparison experiments by removing the proposed query homogeneous GAT, document homogeneous GAT, and neighbor interaction module respectively. The results are presented in Table~\ref{tab:ablation LL PPL} and~\ref{tab:ablation NDCG}. When we remove the query or document homogeneous GAT, the performance degrades on three dataset for both two tasks, which suggests that graph neural networks can extract the collaborative, reformulation and similarity information in the pre-constructed homogeneous graphs to help user behavior prediction. Likewise, the performance drops for both click prediction and relevance estimation tasks on three datasets when we remove the neighbor interaction module, which validates the effectiveness of applying neighbor interactions to capture local and global structural information~\cite{qu2019end}. 

% , from which we obtain the following observations:

% \begin{itemize}[leftmargin=18pt]
%     \item[(1)] The performance drops for both click prediction and relevance estimation tasks on three datasets when we remove the neighbor interaction module, which validates the effectiveness of applying neighbor interactions to capture local and global structural information~\cite{qu2019end}. 
%     \item[(2)] When we remove the query or document homogeneous GAT, the trends in performance for click prediction task are inconsistent for three datasets. The LL and PPL performance degrades on Yandex and TREC2014, but improves on TianGong-ST. However, the NDCG performance for relevance estimation drops dramatically when GATs are removed. This indicates the necessity of GAT modules and neighbor aggregation techniques to improve the performance of click models.
% \end{itemize}

\begin{table}[t]
	\centering
	\caption{The comparison of LL and PPL w.r.t. homogeneous GATs and neighbor interaction module. We perform the following operations respectively: \romannumeral1. remove the query homogeneous GAT; \romannumeral2. remove the document homogeneous GAT; \romannumeral3. remove the neighbor interaction module. The best results are given in bold. $*$ indicates statistically significant improvement (measured by t-test) with p-value $<$ 0.001.}
	\label{tab:ablation LL PPL}
	\resizebox{0.47\textwidth}{!}{
	\renewcommand\arraystretch{1.3}
	\begin{tabular}{c|c c|c c|c c}
	    \toprule
		\hline
		\multicolumn{1}{c|}{\multirow{2}{*}{Model}} &  \multicolumn{2}{c|}{\textbf{Yandex}} &  \multicolumn{2}{c|}{\textbf{TREC2014}} &  \multicolumn{2}{c}{\textbf{TianGong-ST}} \\
		\cline{2-7}
		\multicolumn{1}{c|}{} & LL & PPL & LL & PPL & LL & PPL \\
		\cline{1-7}
		\multicolumn{1}{c|}{0. GraphCM} & $\textbf{-0.2192}^{*}$ & $\textbf{1.2652}^{*}$ & $\textbf{-0.1529}^{*}$ & $\textbf{1.1721}^{*}$ & $\textbf{-0.1629}^{*}$ & $\textbf{1.1938}^{*}$ \\
		\multicolumn{1}{c|}{\romannumeral1. w/o Q.GAT} & -0.2201 & 1.2667 & -0.1601 & 1.1818 & -0.1656 & 1.1968 \\
		\multicolumn{1}{c|}{\romannumeral2. w/o D.GAT} & -0.2198 & 1.2672 & -0.1532 & 1.1738 & -0.1637 & 1.1943 \\
		\multicolumn{1}{c|}{\romannumeral3. w/o Neigh.} & -0.2304 & 1.2755 & -0.1608 & 1.1801 & -0.1702 & 1.2031 \\
 		\hline
 		\bottomrule
	\end{tabular}
	}
\end{table}

\begin{table}[t]
	\centering
	\caption{The comparison of NDCG performance w.r.t. homogeneous GATs and neighbor interaction module for TianGong-ST dataset. The performed operations stay the same as Table~\ref{tab:ablation LL PPL}. The best results are given in bold. $*$ indicates statistically significant improvement (measured by t-test) with p-value $<$ 0.001.}
	\label{tab:ablation NDCG}
	\resizebox{0.475\textwidth}{!}{
	\renewcommand\arraystretch{1.15}
	\begin{tabular}{c c c c c}
	    \toprule
		Model & NDCG@1 & NDCG@3 & NDCG@5 & NDCG@10 \\
		\midrule
		0. GraphCM & $\textbf{0.7388}^{*}$ & $\textbf{0.7189}^{*}$ & $\textbf{0.7466}^{*}$ & $\textbf{0.8671}^{*}$ \\
		\romannumeral1. w/o Q.GAT & 0.7197 & 0.7092 & 0.7351 & 0.8632 \\
		\romannumeral2. w/o D.GAT & 0.6985 & 0.7045 & 0.7323 & 0.8593 \\
		\romannumeral3.w/o Neigh. & 0.7209 & 0.7104 & 0.7382 & 0.8642 \\
		\bottomrule
	\end{tabular}
	}
\end{table}
\section{Conclusion}
In this work, we propose a novel graph-enhanced click model (GraphCM) for web search. We separately model the attractiveness estimation and examination predictor, and apply graph neural networks and neighbor interaction techniques to exploit intra-session and inter-session information for user behavior prediction. Extensive experiments are conducted on three real-world session datasets, which validates the effectiveness of our solution in addressing the data sparsity and cold-start problems. For the future work of research, a promising direction is extending neighbor interactions to higher-orders. Furthermore, we will utilize GraphCM in the downstream tasks (e.g., offline evaluation and optimization).

\section*{Acknowledgement}
The corresponding author Weinan Zhang is supported by ``New Generation of AI 2030'' Major Project (2018AAA0100900) and National Natural Science Foundation of China (62076161, 61772333, 61632017). The work is also sponsored by Huawei Innovation Research Program. We thank Student Innovation Center at Shanghai Jiao Tong University for the provision of GPU computing resources. We thank MindSpore~\cite{mindspore} for the partial support of this work, which is a new deep learning computing framework.

\bibliographystyle{ACM-Reference-Format}
\bibliography{acmart}

\end{document}